\documentclass[aip,cha,reprint,amsmath,amsfonts]{revtex4-1}

\usepackage{bm}
\usepackage{graphicx}
\usepackage{enumerate}

\newtheorem{theorem}{Theorem}

\newcommand{\refcitenosup}[1]{\setcitestyle{numbers}{Ref.~\cite{#1}}\setcitestyle{super}}

\begin{document}

\title{Invariant manifolds and the geometry of front propagation in
  fluid flows}

\author{Kevin A.~Mitchell}
\email{kmitchell@ucmerced.edu}

\author{John R.~Mahoney}
\email{jmahoney3@ucmerced.edu}

\affiliation{School of Natural Sciences, University of California,
Merced, California, 95344}

\date{\today}

\begin{abstract}
  Recent theoretical and experimental work has demonstrated the
  existence of one-sided, invariant barriers to the propagation of
  reaction-diffusion fronts in quasi-two-dimensional
  periodically-driven fluid flows.  These barriers were called burning
  invariant manifolds (BIMs).  We provide a detailed theoretical
  analysis of BIMs, providing criteria for their existence, a
  classification of their stability, a formalization of their barrier
  property, and mechanisms by which the barriers can be circumvented.
  This analysis assumes the sharp front limit and negligible feedback
  of the front on the fluid velocity.  A low-dimensional dynamical
  systems analysis provides the core of our results.
 \end{abstract} 

\maketitle

\begin{quotation}
  The passive advection of particles in a two-dimensional fluid flow
  is a well-studied problem important both for its direct relevance to
  fluid mixing and for its broader connection to Hamiltonian dynamics.
  The latter finds applications ranging from chaotic ionization in
  atomic
  physics~\cite{Bayfield74,Koch95,Mitchell04a,Mitchell04b,Burke09,Burke11,Blumel97}
  to the motion of asteroids in the solar system~\cite{Jaffe02}.  A
  key observation in passive chaotic advection is that particular
  curves---the invariant manifolds---form barriers to particle
  transport~\cite{MacKay84,Ottino89,Rom-Kedar90b,Wiggins92}.  The question
  remains to what extent an \emph{active} material within a fluid flow
  is subject to such barriers.  We are particularly concerned here
  with active systems that generate fronts, e.g., chemical
  reactions~\cite{Tel05,Mezic07,Neufeld09} or sound waves in moving fluids,
  plankton blooms~\cite{Neufeld09,Scotti07,Sandulescu08} in ocean flows, optical
  pulses in an evolving optical medium, phase transitions in liquid
  crystals, or even the spreading of disease within a mixing
  population.  Here, we adopt a model for the propagation of points
  along the front in which each point is viewed as an independent
  active ``swimming'' rod---while the rod always swims perpendicular
  to its orientation, the fluid current both translates and reorients
  the rod.  The invariant manifolds that naturally arise for this
  dynamics are a modification of the traditional manifolds governing
  passive advection.  They prove to be fundamental, one-sided barriers
  to front propagation and strongly influence the patterns of front
  evolution.
\end{quotation}

\section{Introduction}

Recently, the existence of robust barriers to the propagation of
reaction fronts in advection-reaction-diffusion systems was
experimentally demonstrated in both time-independent and time-periodic
vortex-dominated flows~\cite{Mahoney12}.  The flows were
magneto-hydrodynamically-generated, quasi-two-dimensional, and
vortex-dominated, on centimeter length scales.  The reaction was a
ferroin-catalyzed, Belouzov-Zhabotinsky~\cite{Boehmer08} reaction.
See the accompanying article by Bargteil and Solomon~\cite{Bargteil12}
for additional experimental evidence in irregular, vortex-dominated
flows.~\setcitestyle{numbers}In
Ref.~\cite{Mahoney12},\setcitestyle{super}~the barriers were
experimentally shown to be both (i) one-sided, preventing front
propagation in one direction, but not the reverse and (ii) invariant,
staying either fixed in the lab frame or periodically varying (for
periodically driven flows).  These barriers therefore play an
important role in pattern formation in such reactive systems; they
also limit, and could potentially be used to control, the average
reaction propagation speed through the fluid.

The explanation for the barriers in\refcitenosup{Mahoney12}was based
on a low-dimensional dynamical systems approach, which interpreted the
experimental barriers as invariant manifolds attached to unstable
periodic orbits.  Hence, these barriers were named \emph{burning
  invariant manifolds} (BIMs), where ``burning''
\cite{footnote1}
%
emphasizes their relevance to reaction-front propagation and
distinguishes them from the well-established invariant manifolds that
act as barriers in the advective transport of \emph{passive}
tracers~\cite{MacKay84,Ottino89,Rom-Kedar90b,Wiggins92} The main objective of
the present paper is to develop a pedagogical and comprehensive
theoretical justification for, and elaboration of, the picture
introduced in\refcitenosup{Mahoney12}.  In particular, we clarify the
physical properties of BIMs as bounding curves and give explicit
criteria (in the time-independent case) for the existence of BIMs in
laboratory flows.

A complete model of a general advection-reaction-diffusion system is
typically built, at least initially, around a partial differential
equation that couples the local chemical dynamics to the local fluid
flow and vice versa.  In general, the reaction
dynamics can profoundly impact the fluid velocity (e.g. in
combustion.)  However, such feedback appears to be small for the
experiments, such as~\setcitestyle{numbers}Refs.~\cite{Mahoney12} and
\cite{Bargteil12},~\setcitestyle{super} motivating this
work~\cite{Paoletti05,Paoletti05b,Paoletti06,Boehmer08,Schwartz08,Pocheau06,Pocheau08}.
Thus, in this paper, we take the fluid velocity field
$\mathbf{u}(\mathbf{r},t)$ to be prescribed at the outset and to be
unaffected by the subsequent chemical dynamics.  Furthermore, we take
the ``geometric optics'' limit~\cite{Embid95} in which the reaction is
sufficiently fast that a sharp front develops between those regions
that are fully reacted and those that are completely unreacted.  These
assumptions lead to a three-dimensional ODE for the propagation of a
point on the front, discussed below.

For passive tracers, it has long been recognized that one-dimensional
invariant manifolds bound advective transport in two-dimensional
flows~\cite{MacKay84,Ottino89,Rom-Kedar90b,Wiggins92}.  For time-independent
flows, these manifolds form separatrices that strictly separate
distinct regions of the fluid (e.g. different vortex cells.)  For
time-periodic flows, the separatrices split into distinct, but
intersecting, stable and unstable manifolds, which form complicated
heteroclinic (or homoclinic) ``tangles''; advection between regions is
governed through lobes formed by the intersecting manifolds.  More
recently, invariant manifold techniques have been generalized to
aperiodic and even turbulent flows using Lagrangian coherent
structures and finite-time Lypanunov
exponents~\cite{Haller00a,Haller00b,Haller01,Haller02,Shadden05,Shadden06}.

Here, we treat front propagation dynamics as a modification of the
above approach for passive advection.  A point along the front is
viewed as an active ``particle'' (or ``swimmer'') that evolves
according to both fluid advection and its own internal locomotion,
fixed at a constant speed $v_0$ (the ``burning'' speed) in the moving
frame of the fluid.  The corresponding dynamical system is thus
three-dimensional, with two dimensions for the $xy$-position within
the fluid and a third dimension $\theta$ for the orientation of the
local front element.  Initially, it may seem that the invariant
manifolds defined for passive advection are irrelevant for reactive
front propagation, since the reaction can burn right through an
advective separatrix.  However---and this is the central point of this
paper---the invariant manifolds of the advection dynamics can be
suitably and naturally modified to the reaction front scenario; these
are the burning invariant manifolds (BIMs).  The BIMs reduce to
traditional advective invariant manifolds when $v_0 = 0$, but
otherwise may differ dramatically from them.

\begin{figure}
\centering
\includegraphics[width=\linewidth]{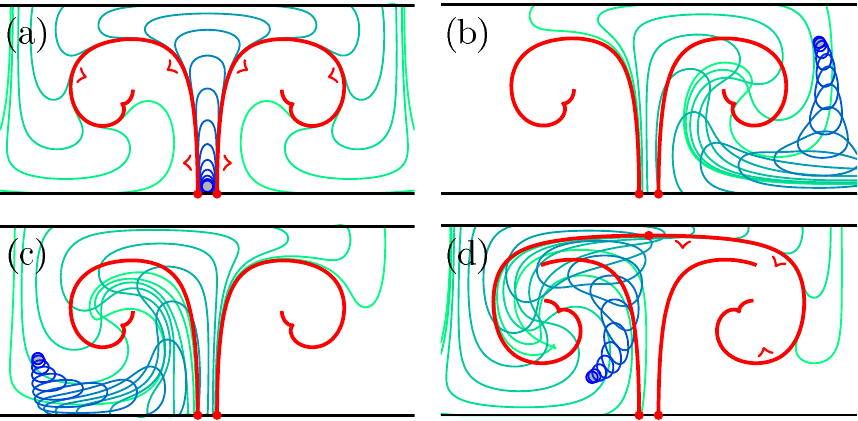}
\caption{(color online) a) Fronts converging (blue to green) upon BIMs
  (bold, red) in a time-independent vortex flow.  Panels b--d show
  fronts initialized at different stimulation points (small open
  circles) burning through BIMs opposite their burning direction but
  bounded by BIMs in their burning direction.  Panel d includes a BIM
  attached to a burning fixed point near the top of the channel, which
  restricts the front in the upper right.}
\label{fig:tindepBIMs}
\end{figure}

\begin{figure}
\centering
\includegraphics[width=\linewidth]{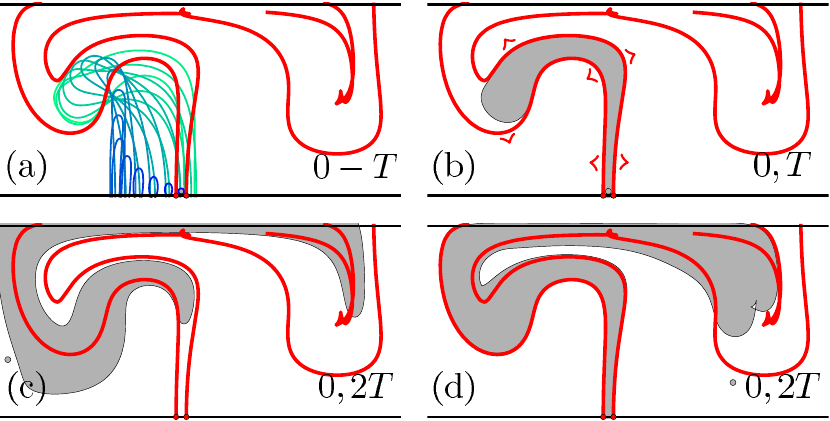}
\caption{(color online) BIMs in a time-periodic flow.  An initial
  stimulation between the BIMs moves outside the BIMs during the
  forcing cycle (panel a) but returns by time $T$ (panel b).  A
  stimulation to the left of the BIM pair evolves through one BIM but
  not the other (panel c).  Similarly for a stimulation on the right
  (panel d).}
\label{fig:tdepBIMs}
\end{figure}

We give a brief overview of the physical role played by BIMs.
Figure~\ref{fig:tindepBIMs} illustrates the BIMs (bold, red) for a
time-independent flow, confined to a horizontal channel with
counter-rotating vortices.  In Fig.~\ref{fig:tindepBIMs}a, the
reaction is catalyzed at the channel bottom between the BIMs, and then
generates the sequence of reaction fronts, forming an upward
propagating plume, ultimately rolled up by the vortices.  The fronts
are bounded by the BIMs on either side, with the arrows normal to the
BIMs denoting the ``burning'' direction of each BIM.  A BIM forms a
barrier to fronts propagating in its burning direction (as in
Fig.~\ref{fig:tindepBIMs}a), but not to fronts impinging in the
opposite direction, as illustrated in Figs.~\ref{fig:tindepBIMs}b--d.
In Figs.~\ref{fig:tindepBIMs}b--d the fronts are catalyzed at
different stimulation points (small open circles) and approach the
BIMs from different directions.  In each case, the fronts pass through
a BIM when approaching opposite the BIM's burning direction but are
stopped when approaching in the burning direction.

In Fig.~\ref{fig:tdepBIMs}, the vortices oscillate periodically in the
horizontal direction.  The BIMs are shown at a single phase of the
driving only.  An initial stimulation between the BIMs in
Fig.~\ref{fig:tdepBIMs}a grows outward in a series of reaction fronts.
During the first period, the front migrates outside the BIM channel in
Fig.~\ref{fig:tdepBIMs}a, but returns to it exactly one driving
period $T$ after initialization (shown by the gray ``burned'' region
in Fig.~\ref{fig:tdepBIMs}b.)  Figures~\ref{fig:tdepBIMs}c and
\ref{fig:tdepBIMs}d show two additional stimulation points, and their
subsequent evolution after two driving periods (in gray).  In both
cases, BIMs bound the front evolution in their burning direction, but
not in the opposite direction.

This paper is organized as follows.  Section~\ref{sec:ODE} introduces
the finite-dimensional dynamical system for a point along the front,
and Sec.~\ref{sec:Compatibility} discusses how extended fronts are
represented within this framework.  Section~\ref{sec:grid} compares
this ODE approach to a grid-based computation.  Front singularities
and collisions are discussed in Sec.~\ref{sec:intersections}.
Section~\ref{sec:nopassing} formalizes the observation that fronts
cannot pass one another, in what we call the ``no-passing'' lemma.
Section~\ref{sec:BFP} derives existence (Sec.~\ref{sec:existence}) and
stability (Sec.~\ref{sec:stability}) results for fixed points of the
front dynamics, focusing on time-independent flows.  Linear flows are
examined in Sec.~\ref{sec:linear}.  The existence of fixed points is
constrained by a topological index theory developed in
Sec.~\ref{sec:topindex}.  The heart of the paper is the introduction
of BIMs in Sec.~\ref{BIMBarriers}, which establishes that BIMs can
(locally) be interpreted as fronts (Sec.~\ref{sec:BIMfronts}) that
provide one-sided barriers to front propagation
(Sec.~\ref{sec:onesided}).  Sections \ref{sec:cusps} and
\ref{sec:movepast} discuss two mechanisms for circumventing these
barriers.  Section~\ref{sec:convergence} describes how BIMs attract
fronts, providing a mechanism to measure BIMs directly in the
laboratory.  Finally, Sec.~\ref{sec:limits} connects features of BIMs
to the geometry of the simpler cases of either pure (passive)
advection or pure (nonadvecting) front propagation.

\section{Front propagation in an advecting medium: fundamental
  concepts}

We analyze the general problem of a front propagating through a
two-dimensional medium flowing with velocity
$\mathbf{u}(\mathbf{r},t)$, specified as a function of position
$\mathbf{r} = (x,y)$ and time $t$.  In the local co-moving frame of
the medium, propagation is assumed to progress normal to the front at
a speed $v_0$ (the burning speed), which is constant, independent of
position, propagation direction, or the local front
curvature~\cite{Neufeld09}.  Finally, the velocity field
$\mathbf{u}(\mathbf{r},t)$ is assumed to be established independently
from the front dynamics itself, i.e. the development of the front does
not feedback to modify $\mathbf{u}$.

Our front propagation theory has been successfully applied to the
driven advection-reaction-diffusion (ARD) systems studied
experimentally by Solomon~\cite{Bargteil12,Mahoney12}.  In these
experiments, the reaction time scale is much faster than the advection
time scale (i.e. large Damk\"{o}hler number), meaning that the front
is thin compared to the scale of the flow---the so-called ``geometric
optics'' limit~\cite{Embid95}.  In this limit, a fluid element is
either fully reacted or unreacted; hence, the distribution of reaction
products within the fluid is entirely determined by the boundary of
the reacted, or ``burned'', region.  Throughout this paper, such ARD
systems form the prototype example, though our theoretical framework
should lend insight into pattern formation in other systems that
combine front propagation and advection, including not only chemical
reaction dynamics, but optics, phase-transitions, and even quantum
phenomena.

\subsection{A three-dimensional ODE for the propagation of
  front elements in a two-dimensional flow}
\label{sec:ODE}

\begin{figure}
\centering
\includegraphics[width=\linewidth]{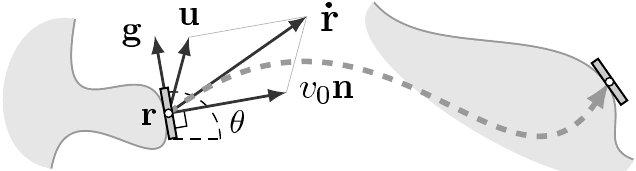}
\caption{Each reaction front element independently propagates forward
  under the ODEs (\ref{r3}) and (\ref{eq:3DODE}).}
\label{fig:3DODE}
\end{figure}

A front element, i.e. a point on the boundary between burned and
unburned regions, is described by both its position $\mathbf{r} =
(x,y)$ and its unit normal $\mathbf{n}$ (Fig.~\ref{fig:3DODE}).
Alternatively, $\mathbf{n}$ can be replaced by the unit tangent
$\mathbf{g}$, defined here by a right-handed turn by $\pi/2$,
\begin{align}
\mathbf{g} = \mathsf{J} \mathbf{n}, 
& \quad \text{where} \quad \mathsf{J}  = 
\left[
\begin{array}{cc}
0 & -1 \\
1 & 0 
\end{array}
\right].
\label{r7}
\end{align}
Under the above assumptions, a front element evolves
via~\cite{Oberlack10}
\begin{subequations}
\begin{align}
\mathbf{\dot{r}} &= \mathbf{u} + v_0
\mathbf{n} \label{r19} \\
& = \mathbf{u} - v_0 \mathsf{J}
\mathbf{g}, \label{r1} \\
\mathbf{\dot{g}} &= 
[\mathbf{g}  \cdot  (\nabla \mathbf{u}) \mathbf{n}]
\mathbf{n} \label{r20} \\
& = \mathbf{g}  \cdot (\nabla \mathbf{u}) 
-  [\mathbf{g}  \cdot (\nabla \mathbf{u}) \mathbf{g}]
\mathbf{g}, 
\label{r2}
\end{align}
\label{r3}
\end{subequations}
where dots denote time derivatives and $\nabla \mathbf{u}$ is the
velocity gradient tensor with components $(\nabla \mathbf{u})_{ij} =
u_{j,i}$, $i,j = x,y$, where a comma denotes partial differentiation.
The equivalence of Eqs.~(\ref{r19}) and (\ref{r1}) and of
Eqs.~(\ref{r20}) and (\ref{r2}) is straightforward, the latter
requiring $|\mathbf{g}|=1$.  To justify Eqs.~(\ref{r19}) and
(\ref{r2}), first consider the nonburning case $v_0 = 0$ for which
front elements should evolve under advection alone.
Equation~(\ref{r19}) reduces to the appropriate advective form
$\mathbf{\dot{r}} = \mathbf{u}$.  Eq.~(\ref{r2}) remains unchanged;
the first term generates the dynamics of advection on the
(unnormalized) tangent vector, whereas the second enforces the
normalization $|\mathbf{g}| = 1$.  [Equation~(\ref{r2}) readily yields
$d | \mathbf{g} |^2 / dt = 2 \mathbf{g} \cdot \mathbf{\dot{g}} = 0$
when $|\mathbf{g}| = 1$.]  Considering the case $v_0 \neq 0$, the only
necessary change to Eqs.~(\ref{r19}) and (\ref{r2}) is to add the
burning velocity $v_0 \mathbf{n}$ normal to the front in
Eq.~(\ref{r19}).  An appealing physical analogy can be obtained by
recognizing that Eq.~(\ref{r3}) also describes the motion of a thin
rod-shaped swimmer aligned with $\mathbf{g}$ and propelling itself
\emph{normal} to its alignment direction at constant speed $v_0$ in
the local fluid frame.

Since $|\mathbf{g}|= 1$ for all time, Eq.~(\ref{r3}) reduces to a
three-dimensional ODE in $(x,y, \theta)$, where $\theta$ is the
orientation angle between $\mathbf{g}$ and the positive $x$-axis
(Fig.~\ref{fig:3DODE}), i.e. $\mathbf{g} = ( \cos \theta, \sin
\theta)$.
\begin{subequations}
\begin{align}
\displaystyle \dot{x} & = u_x + v_0 \sin{\theta},  \\
\dot{y} &= u_y - v_0 \cos{\theta},  \\
\displaystyle \dot{\theta} &= - u_{x,x} \sin{\theta} \cos{\theta}
- u_{x,y} \sin^2{\theta} + \nonumber \\
& \quad u_{y,x} \cos^2{\theta} + u_{y,y} \sin{\theta} \cos{\theta}.
\label{r18}
\end{align}
\label{eq:3DODE}
\end{subequations}
For incompressible fluids, $u_{y,y} = -u_{x,x}$, simplifying
Eq.~(\ref{r18}).  We emphasize that even though Eq.~(\ref{eq:3DODE})
is three-dimensional, the fluid flow itself is \emph{two}-dimensional.
The reduction from four [Eq.~(\ref{r3})] to three
[Eq.~(\ref{eq:3DODE})] dimensions is useful for both numerical
computations and many theoretical arguments.

\subsection{Front compatibility criterion}
\label{sec:Compatibility}

A front may either be viewed as a curve in the two-dimensional
$xy$-space or in the three-dimensional $xy\theta$-space (or indeed
even in the four-dimensional $\mathbf{r g}$-space.)  We clarify the
relationship between these two representations here.  First, we
parameterize fronts by their Euclidean length $\lambda$, measured
along the front in $xy$-space.  By convention $\lambda$ increases in
the $+\mathbf{g}$ direction; this is equivalent to keeping the burned
region on your left while traversing the curve.  Then any
(differentiable) curve $\mathbf{r}(\lambda)$ representing a front in
$xy$-space lifts to a unique curve
$(\mathbf{r}(\lambda),\theta(\lambda))$ in $xy\theta$-space, where
$\theta(\lambda)$ is the unique (modulo $2 \pi$) solution to
\begin{equation}
  ( \cos \theta(\lambda), \sin \theta(\lambda)) 
=  d \mathbf{r} / d \lambda.  
\label{r4}
\end{equation}
On the other hand, a general curve $(\mathbf{r}(\lambda),
\theta(\lambda))$ in $xy\theta$-space is typically not the lift of a
curve in $xy$-space---\emph{only} those curves in $xy\theta$-space
satisfying Eq.~(\ref{r4}) represent fronts.  We thus call
Eq.~(\ref{r4}) the \emph{front compatibility criterion} for a curve
$(\mathbf{r}(\lambda), \theta(\lambda))$.  Geometrically, it is the
requirement that at every point $\lambda$ of the curve, the
two-dimensional vector $d \mathbf{r} /d \lambda$ points in the
direction $\mathbf{g}(\lambda)$, \emph{defined} by
$\mathbf{g}(\lambda) = ( \cos \theta(\lambda), \sin \theta(\lambda))$.
Thus the front compatibility criterion is equivalently written
\begin{equation}
\mathbf{g} \cdot \mathsf{J} \frac{  d \mathbf{r} }{d \lambda} =
\mathbf{n} \cdot \frac{  d \mathbf{r} }{d \lambda} = 0.
\label{r6}
\end{equation}
For curves in $xy\theta$-space, we reserve the term \emph{front} for
those curves satisfying the front compatibility criterion.  To
summarize, a front is represented either by a parameterized curve in
$xy$-space or a parameterized curve in $xy\theta$-space that satisfies
the front compatibility criterion, and there is a one-to-one
correspondence between these representations.

If the front compatibility criterion [Eq.~(\ref{r6})] holds along a
curve $(\mathbf{r}(\lambda,t),\theta(\lambda,t))$ at some initial
time, then it must hold for all future times as the curve evolves.
Physically, this must be the case, since the time-evolution of a front
must remain a front.  To check that it follows from Eq.~(\ref{r3}),
assume first that Eq.~(\ref{r6}) holds at an initial time.  Then,
\begin{align}
\frac{d}{dt}&\left(\mathbf{g}  \cdot \mathsf{J} \frac{  d \mathbf{r} }{d
  \lambda}\right) \nonumber \\
&= 
[\mathbf{g}  \cdot (\nabla \mathbf{u}) \mathbf{n}]
\mathbf{n}  \cdot \mathsf{J} \frac{  d \mathbf{r} }{d
  \lambda} 
+  \mathbf{g} \cdot  \mathsf{J} \left[ (\nabla \mathbf{u})^T \frac{d \mathbf{r}}{d\lambda} - v_0 \mathsf{J}
\frac{d \mathbf{g}}{d\lambda} \right] \nonumber \\
&=
-\mathbf{g}  \cdot (\nabla \mathbf{u}) \mathbf{n}
+ \mathbf{n} \cdot  (\nabla \mathbf{u})^T \mathbf{g} = 0,
\end{align}
where the first equality follows from Eq.~(\ref{r3}) and the second
from the fact that $\mathbf{g} \cdot (d\mathbf{g}/d\lambda) = 0$ and
Eqs.~(\ref{r7}) and (\ref{r4}).  Thus, if Eq.~(\ref{r6}) holds at an
initial time, it holds for all future times.

\subsection{Example Fluid Flow}

Equations~(\ref{r3}) and (\ref{eq:3DODE}) are valid for an arbitrary
flow $\mathbf{u}(\mathbf{r},t)$.  For illustrative purposes (see
Figs.~\ref{fig:tindepBIMs}, \ref{fig:tdepBIMs}, \ref{fig:sims},
\ref{fig:cusp}, and \ref{fig:evolution}), we use the following
incompressible velocity field
\begin{equation}
\begin{aligned}
u_x(x, y, t) &= +\sin(\pi[x + b \sin (\omega t)]) \cos(\pi y ), \\
u_y(x, y, t) &= -\cos(\pi[x + b \sin (\omega t)]) \sin(\pi y ). 
\end{aligned}
\label{eq:velfield} 
\end{equation}
This flow represents an array of alternating vortices, periodically
driven with frequency $\omega$ and amplitude $b$.  It has been
extensively used to successfully model driven laboratory
flows~\cite{solomon88,camassa91,Cencini03,Paoletti05,Paoletti05b,
  Mahoney12,Bargteil12}
\cite{footnote2}.

\subsection{Two approaches to front propagation: curve-based versus
  grid-based}

\label{sec:grid}

\begin{figure}
\includegraphics[width=1\linewidth]{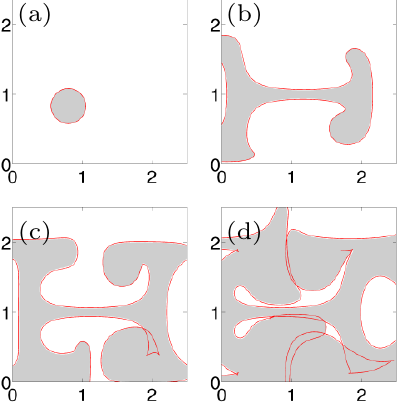}
\caption{a) An initial circular burned region at $t = 0.25 T$, with
  the black front surrounding it; Panels b--d show the forward
  evolution at times $t = 0.69T, 0.97T, 1.16T$.  The black front is
  evolved using Eq.~(\ref{eq:3DODE}) with $v_0 = 0.2$ and $\mathbf{u}$
  given by Eq.~(\ref{eq:velfield}), with $b = 0.25$, and $\omega = 2$.
  The gray burned domain is evolved according to a grid-based,
  split-step algorithm, using 1668x1668 grid points within the
  $xy$-domain shown (full integration domain is larger), with time
  step $\Delta t = 2\pi/(32 \omega)$.  }
\label{fig:sims}
\end{figure}

Under the assumptions laid out in Sec.~\ref{sec:ODE}, an entire front
can be evolved by independently evolving individual front elements.
Numerically, this requires a sufficient density of points along the
front to resolve it to a desired accuracy.  As the front evolves, the
separation between some of the points typically increases, requiring
the addition of new interpolated points to maintain sufficient
density.  We use an algorithm similar to 
\setcitestyle{numbers}
Ref.~\cite{You91}.
\setcitestyle{super}

\setcitestyle{numbers}Refs.~\cite{Abel01,Abel02,Cencini03}\setcitestyle{super}
used a Euclidean-grid-based approach to evolve the entire burned
region itself.  In such an approach, each grid point is recorded as
either burned or unburned at a given time.  A split-step scheme,
consisting of a pure burning step and a pure advection step, is used
to update each gridpoint in time.  (For details, see
\setcitestyle{numbers}
Refs.~\cite{Abel01,Abel02,Cencini03}.)\setcitestyle{super} In contrast,
our approach focuses solely on the front itself.

Fig.~\ref{fig:sims} compares the above two algorithms for front
evolution.  Fig.~\ref{fig:sims}a shows an initial burned circular
region.  The black curve denotes the front $( x(\lambda), y(\lambda),
\theta(\lambda))$ that forms the initial condition (at $t = 0.25 T$,
with $T = 2\pi/\omega$) for evolution via Eqs.~(\ref{eq:3DODE}) and
(\ref{eq:velfield}).  A corresponding grid-based computation is also
performed, with initial condition consisting of the burned grid points
inside the circle (shaded gray.)  Fig.~\ref{fig:sims}b shows the
independent evolution of both the black circle [via
Eq.~(\ref{eq:3DODE})] and the gray grid points [via the split-step
algorithm] up to $t = 0.69 T$.  The two techniques agree well, with
the black curve marking the boundary of the gray region, thereby
providing added numerical confirmation for the validity of
Eq.~(\ref{eq:3DODE}).  The tiny visible discrepancy is due to the
finite grid resolution.

\subsection{Front intersections and cusps}

\label{sec:intersections}

Fig.~\ref{fig:sims}c continues the evolution to $t = 0.97 T$.  Part of
the evolved front now lies within the burned (gray) region.  While we
refer to the entire evolved curve (either in $xy$-space or
$xy\theta$-space) as the \emph{front}, we refer to those segments of
the front separating the burned from unburned regions as the
\emph{bounding} front; those intervals of the front not in the
bounding front thus lie interior to the burned region.  Note that the
bounding front need not be connected, but can have disconnected
segments defining voids, as seen at time $t = 1.16T$
(Fig.~\ref{fig:sims}d).  Figures~\ref{fig:sims}c and \ref{fig:sims}d
continue the excellent agreement between the evolution algorithms.
Note that even though the front may intersect itself when projected
into the $xy$-plane, it cannot self intersect in the full
$xy\theta$-space, as mandated by the uniqueness theorem for solutions
of ODEs.

Figures \ref{fig:sims}c and \ref{fig:sims}d illustrate two mechanisms
by which a front may penetrate the interior of the burned region.  In
the transition from Fig.~\ref{fig:sims}b to Fig.~\ref{fig:sims}c, a
so-called swallowtail catastrophe is formed by the $xy$-projection of
the front~\cite{Arnold83,Arnold89}; the characteristic swallowtail
shape, with its two cusps, is seen interior to the burned region in
Fig.~\ref{fig:sims}c.  Despite singularities in its $xy$-projection,
the front in $xy\theta$-space remains smooth throughout.  In the
transition from Fig.~\ref{fig:sims}c to Fig.~\ref{fig:sims}d, fronts
penetrate the interior through the head-on collision of distinct
locations along the front.  Both mechanisms for producing front
intersections in the $xy$-plane are characteristic of front evolution
in optics~\cite{Arnold83,Arnold89}, for example, and do not require
the presence of advection.

\subsection{Front no-passing lemma}
\label{sec:nopassing}

Consider two fronts, with one front trailing the other.  Since both
fronts have the same speed in the local fluid frame, it is clear that
the trailing front can not catch up to and pass the leading front.
This no-passing result will be central to understanding why BIMs form
one-sided barriers to front evolution in Sec.~\ref{BIMBarriers}.

\begin{figure}
\centering
\includegraphics[width=\linewidth]{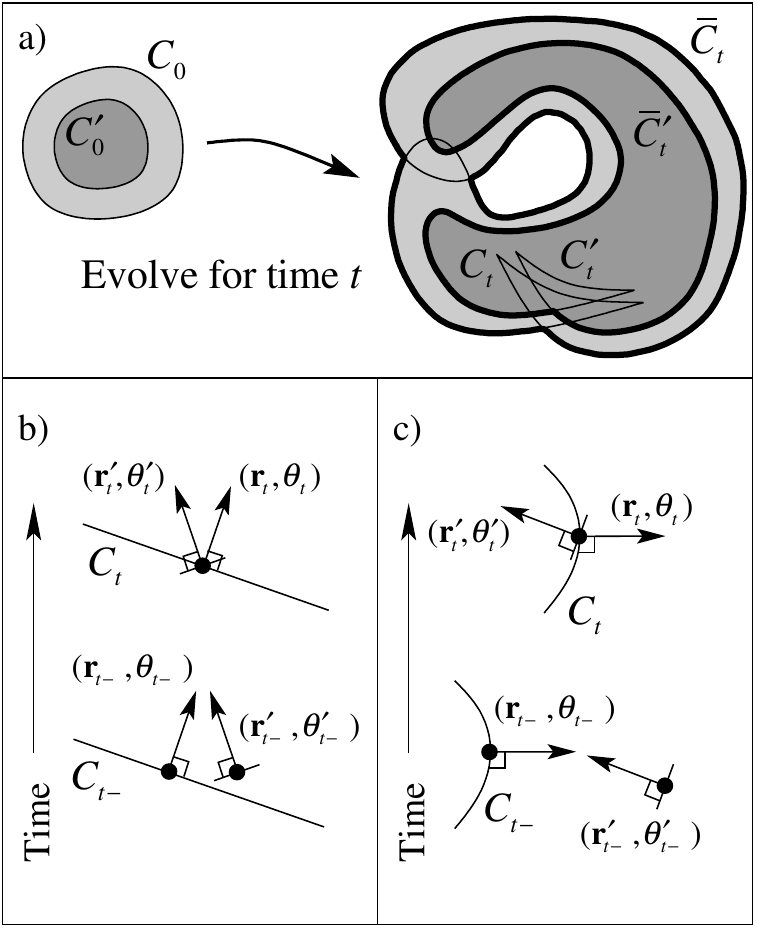}
\caption{a) The fronts $C_0$ and $C_0'$ evolve into $C_t$ and $C_t'$.
  The bounding fronts $\bar{C}_t$ and $\bar{C}_t'$ (in bold) have the
  same nested ordering as $C_0$ and $C_0'$. Panels b and c illustrate
  a front element $(\mathbf{r}_t', \theta')$ (of $C_t'$, not shown)
  approaching $C_t$ from the front and colliding with it.}
\label{fig:nopassing}
\end{figure}

The intuitive no-passing statement above can be recast more precisely
as follows.  Let $C_0$ and $C_0'$ be the leading and trailing fronts,
and assume that they are represented by (piecewise smooth)
nonintersecting closed loops in $xy$-space, with $C_0'$ nested inside
$C_0$ (Fig.~\ref{fig:nopassing}a).  Without loss of generality, we
assume that both are burning outward, meaning that their interior
domains are burned.  (The case of both burning inward, with their
interior domains unburned, can be handled analogously.)  The fronts
may extend to infinity or coincide with a boundary wall for part of
their length, so long as the burned region of $C_0'$ is entirely
included within the burned region of $C_0$.  As $C_0$ and $C_0'$
evolve forward for some time $t$, according to Eq.~(\ref{eq:3DODE}),
they produce new fronts $C_t$ and $C_t'$.  In general, $C_t$ and
$C_t'$ might self-intersect and intersect one another (in the
$xy$-plane), as shown in Fig.~\ref{fig:nopassing}a.  However, despite
these intersections, the bounding fronts $\bar{C}_t$ and $\bar{C}_t'$
(in bold) will not intersect and will retain their original nested
ordering, with $\bar{C}_t'$ lying within the burned region defined by
$\bar{C}_t$.  We call this the no-passing lemma.  It follows from
the simple physical observation that the innermost burned region must
always remain a subset of the outermost burned region.

The no-passing lemma can be recast as a local formulation in terms of
non-closed segments of the fronts $C_t$ and $C_t'$.  Suppose
$(\mathbf{r}_t',\theta'_t)$ is a front element (of curve $C_t'$) whose
position $\mathbf{r}_t'$ coincides with a point $\mathbf{r}_t$ on the
front $C_t$, but whose $\theta'_t$ does not agree with the local
orientation of $C_t$.  Then a short time $\epsilon$ earlier, the front
element $(\mathbf{r}_{t-\epsilon}',\theta'_{t-\epsilon})$ must lie in
front of $C_{t-\epsilon}$ (Figs.~\ref{fig:nopassing}b and
\ref{fig:nopassing}c).  Said another way, the front element
$(\mathbf{r}_t',\theta'_t)$ can not catch up to and pass $C_t$ from
behind.  Of course, there is nothing preventing
$(\mathbf{r}_t',\theta'_t)$ from colliding with $C_t$ from the front,
as in Figs.~\ref{fig:nopassing}b and \ref{fig:nopassing}c.  Formally,
the local no-passing formulation may be recast as the inequality
\begin{equation}
\dot{\mathbf{r}}' \cdot \mathbf{n} <
\dot{\mathbf{r}} \cdot \mathbf{n},
\label{r21}
\end{equation}
where $\dot{\mathbf{r}}'$ and $\dot{\mathbf{r}}$ are the velocities of
the two front elements at the time their $xy$-positions coincide and
$\mathbf{n}$ is the unit normal to $C_t$.  Equation~(\ref{r21})
follows directly from $\dot{\mathbf{r}} = \mathbf{u}(\mathbf{r}) + v_0
\mathbf{n}$ [Eq.~(\ref{r3}a)], $\dot{\mathbf{r}}' =
\mathbf{u}(\mathbf{r}) + v_0 \mathbf{n}'$, and $\mathbf{n} \cdot
\mathbf{n'} < 1$.  Finally, note that the roles of
$(\mathbf{r}_t',\theta'_t)$ and $(\mathbf{r}_t,\theta_t)$ can be
reversed, so that $(\mathbf{r}_t,\theta_t)$ can not catch up to and
pass $C_t'$ either; mathematically, $\dot{\mathbf{r}} \cdot
\mathbf{n}' < \dot{\mathbf{r}}' \cdot \mathbf{n}'.$

\section{Burning fixed points}

\label{sec:BFP}

By a \emph{burning fixed point}, we mean either an equilibrium point
of the differential equation (\ref{r3}) [or (\ref{eq:3DODE})], for
time-independent flows, or a fixed point of the corresponding
Poincar\'{e} map, for time-periodic flows.  For the remainder of
Sec.~\ref{sec:BFP}, we restrict our focus to time-independent flows,
for which we give a precise characterization of the existence and
stability of burning fixed points.

\subsection{Existence criteria for burning fixed points in
  time-independent flows}

\label{sec:existence}

At a burning fixed point in a time-independent flow, Eq.~(\ref{r19})
implies $v_0 \mathbf{n} = -\mathbf{u}$, so that the fluid velocity is
exactly balanced by the burning velocity.  Eq.~(\ref{r20}) on the
other hand implies $0 = \mathbf{n} \cdot \mathbf{\dot{g}} = \mathbf{g}
\cdot (\nabla \mathbf{u}) \mathbf{n}$.

\begin{theorem}
  A burning fixed point $(\mathbf{r},\mathbf{g})$ occurs if and
  only if the following two conditions are met.
\begin{alignat}{3}
 {}& (i) \quad & v_0 \mathbf{n} &= -\mathbf{u}, \label{r5} \\
 {}& (ii) \quad & \mathbf{g} \cdot (\nabla \mathbf{u}) \mathbf{n} &= 0.
  \label{r8}
\end{alignat}
Condition (ii) is equivalent to
\begin{equation}
(ii') \quad (\nabla \mathbf{u})^T \mathbf{g} = \mu  \mathbf{g},
\label{r9}
\end{equation}
for some (necessarily real) eigenvalue $\mu$.
\end{theorem}

\begin{figure}
\centering
\includegraphics[width=\linewidth]{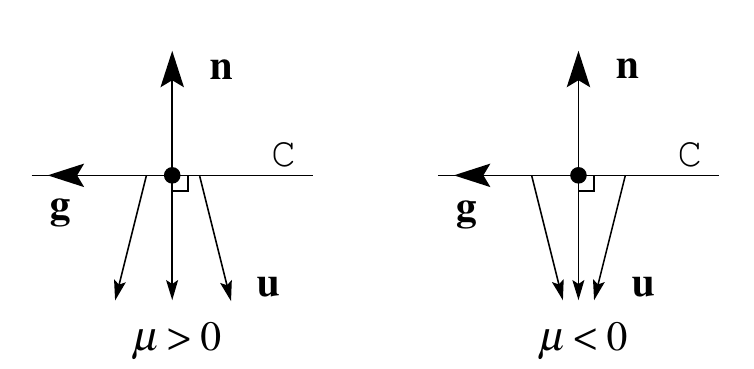}
\caption{Illustration of flows with $\mu > 0$ and $\mu<0$ about a
  burning fixed point, at which $\mathbf{u}$ is normal to the set
  $\mathcal{C}$ defined by $|\mathbf{u}| = v_0$.  The burning
  direction $\mathbf{n}$ points opposite $\mathbf{u}$.}
\label{fig:ConDiFields}
\end{figure}

Together, Eqs.~(\ref{r5}) and (\ref{r8}) imply $\mathbf{g} \cdot
\nabla | \mathbf{u} |^2 = 0,$ i.e. $ \mathbf{g}$ is tangent to the
level set $|\mathbf{u}| = v_0$ (assuming $\nabla | \mathbf{u} |^2 \ne
0$).  Alternatively, $\mathbf{n}$ is normal to the level set, so that
by Eq.~(\ref{r5}) $\mathbf{u}$ is also normal to the level set.

\begin{theorem}
\label{t2}
Define the level set $\mathcal{C}$ in $xy$-space by $|\mathbf{u} | =
v_0$.  Assuming $\nabla | \mathbf{u} |^2 \ne 0$ on $\mathcal{C}$, the
set of burning fixed points consists of exactly those points
$(\mathbf{r},\mathbf{g})$ at which $\mathbf{r}$ is on $\mathcal{C}$
and $\mathbf{u}(\mathbf{r})$ points normal to $\mathcal{C}$.  The
burning direction $\mathbf{n} = -\mathsf{J}\mathbf{g}$ points opposite
$\mathbf{u}$.
\end{theorem}
Figure~\ref{fig:ConDiFields} illustrates Theorem~\ref{t2} for $\mu > 0$
and $\mu < 0$. 

\subsection{Burning fixed points for linear fluid flows}

\label{sec:linear}

The case of linear flows lends insight into the existence and local
structure of burning fixed points.

\subsubsection{Hyperbolic flows}
\label{sec:hyperbolic}

The hyperbolic flow
\begin{equation}
u_x = -Ax, \quad u_y = +Ay,
\label{r22}
\end{equation}
with strength $A > 0$, has a unique hyperbolic advective fixed point
at the origin.  This flow generates the burning dynamics
\begin{subequations}
\begin{align}
\dot{x} &= -Ax + v_0 \sin \theta,\\
\dot{y} &= +Ay - v_0 \cos \theta,\\
\dot{\theta} &= +2A \cos \theta \sin \theta.
\end{align}
\label{r10}
\end{subequations}
Thus, any fixed point $(x_*, y_*, \theta_*)$ must have $\theta_* = 0,
\pi, \pi/2, 3\pi/2$.  Each $\theta_*$ then yields a unique value of
$x_*$ and $y_*$, summarized in Table~\ref{Table:BFP}.  The advective
hyperbolic fixed point thus produces four burning fixed points as
$v_0$ deviates from 0.

\begin{table}
\begin{tabular}{l|c|l|c|c}
 & BFP $(x_*, y_*, \theta_*)$ & Eigenvalues& Stability & 1D BIM dir\\
\hline
1 & $(0, v_0/A, 0)$ & $-A, +A, +2A$ & SUU & $\mathbf{\hat{x}}$\\
2 & $(0, -v_0/A, \pi)$ & $-A, +A, +2A$ & SUU &$\mathbf{\hat{x}}$\\ 
3 & $(v_0/A, 0, \pi/2)$ & $-A, +A, -2A$ & SSU &$\mathbf{\hat{y}}$\\
4 & $(-v_0/A, 0, 3\pi/2)$ & $-A, +A, -2A$ & SSU &$\mathbf{\hat{y}}$
\end{tabular}
\caption{\label{Table:BFP} Burning fixed points for 
  hyperbolic flow Eq.~(\ref{r22}).}
\end{table}

\begin{figure}
  \includegraphics[width=\linewidth]{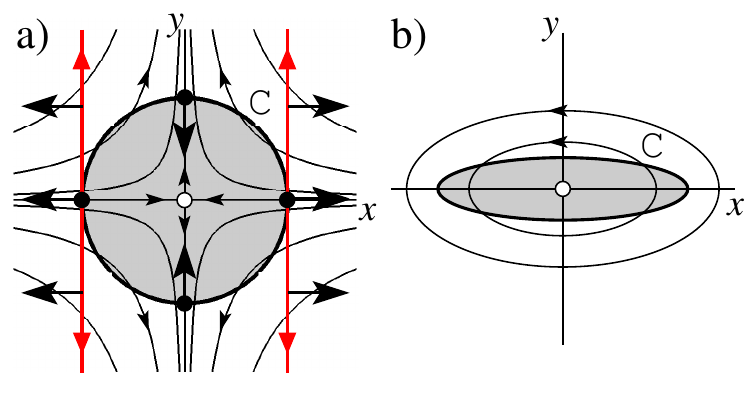}
  \caption{ (color online) a) The four burning fixed points (filled
    dots) in a hyperbolic flow lie on the circle $\mathcal{C}$ (the
    level set $|\mathbf{u}| = v_0$) which surrounds the advective
    fixed point (open dot).  The shaded region satisfies $|\mathbf{u}|
    < v_0$.  The burning velocities (large arrows) of these four
    points are normal to the circle, pointing opposite the flow.  The
    left and right points have stability SSU, and the vertical lines
    passing through them are their unstable BIMs.  The triangular
    (unbarbed) arrows denote the unstable direction along the BIMs.
    b) An elliptic flow has no burning fixed points since the flow
    lines are never normal to the ellipse $\mathcal{C}$ bounding the
    shaded region.}
\label{fig:hyperbolic}
\end{figure}

Figure~\ref{fig:hyperbolic}a shows the four burning fixed points.
Consistent with Theorem~\ref{t2}, these four points lie on the circle
$A |\mathbf{r}| = |\mathbf{u}| = v_0$, bounding the shaded region
$|\mathbf{u}| < v_0$.  Furthermore, these are exactly the four points
where the hyperbolic flow is normal to the circle.

The linearization of Eq.~(\ref{r10}) yields
\begin{equation}
\mathsf{L} = 
\frac{\partial (\dot{x}, \dot{y}, \dot{\theta})}
{\partial  (x, y, \theta)}=
\left(
\begin{array}{ccc}
-A & 0 & v_0 \cos \theta \\
  0 & A & v_0 \sin \theta \\
  0 & 0 & 2A \cos 2\theta 
\end{array}
\right).
\end{equation}
The eigenvalues of $\mathsf{L}$ at the burning fixed points are
$(-A,A,\pm 2A)$, summarized in Table~\ref{Table:BFP}.  Each fixed
point is hyperbolic with either stability stable-stable-unstable (SSU)
or stable-unstable-unstable (SUU).  Each fixed point with SSU (SUU)
stability thus has a one-dimensional unstable (stable) invariant
manifold, which is a burning invariant manifold (BIM).  These
manifolds have eigendirections $(\mathbf{\hat{y}},0)$ (for SSU) or
$(\mathbf{\hat{x}},0)$ (for SUU).  In both cases, the $xy$-projection
of the eigendirection is aligned with the orientation $\mathbf{g}$ of
the fixed point.

Focusing on the two burning fixed points on the $x$-axis, their
unstable BIMs are the straight lines $\{ (\pm v_0/A, y, \pi/2) |
-\infty < y < \infty \}$, which project to the two vertical lines in
Fig.~\ref{fig:hyperbolic}a.  Note that these BIMs satisfy the front
compatibility condition Eq.~(\ref{r4}), a point we return to in
Sec.~\ref{sec:BIMfronts}.

\subsubsection{Elliptic flows}

The elliptic flow
\begin{equation}
u_x = y/A, \quad u_y = -Ax,
\end{equation}
with $A>0$, has a unique elliptic advective fixed point at the origin.
It generates the burning dynamics,
\begin{subequations}
\begin{align}
\dot{x} &= y/A + v_0 \sin \theta,\\
\dot{y} &= -Ax - v_0 \cos \theta,\\
\dot{\theta} &= -(\sin^2 \theta)/A - A \cos^2 \theta ,
\label{r11}
\end{align}
\end{subequations}
from which we see that $\dot{\theta} = 0$ has no solutions, and hence
there are no burning fixed points in a purely elliptic flow.  This can
also be seen as a consequence of Theorem~\ref{t2}, as illustrated in
Fig.~\ref{fig:hyperbolic}b.  Any burning fixed point must lie on the
elliptic level set $[ (Ax)^2 + (y/A)^2 ]^{1/2} = |\mathbf{u}| = v_0$
bounding the shaded region.  However, the flow is never normal to this
set, and hence there are no burning fixed points.

\subsection{Stability of burning fixed points for general
  time-independent flows}

\label{sec:stability}

For the main results on the stability of burning fixed points, the
reader may skip ahead to Theorems~\ref{t3} and \ref{t4}.  The
linearization of Eq.~(\ref{r3}) yields the following four-by-four
matrix, decomposed into two-by-two blocks.

\begin{equation}
\mathsf{X} = 
\left(
\begin{array}{c|c}
\partial \mathbf{\dot{r}} / \partial \mathbf{r} &  
\partial \mathbf{\dot{r}} / \partial \mathbf{g} \\
\hline 
\partial \mathbf{\dot{g}} / \partial \mathbf{r} &  
\partial \mathbf{\dot{g}} / \partial \mathbf{g} 
\end{array}
\right)
=
\left(
\begin{array}{c|c}
(\nabla \mathbf{u})^T &  
-v_0 \mathsf{J}  \\
\hline
\mathbf{n} \otimes \mathbf{a}  &  
\mathsf{D}
\end{array}
\right),
\end{equation}
where
\begin{align}
\mathsf{D} &= 
(\mathbf{g} \otimes \mathbf{n} + \mathbf{n} \otimes \mathbf{g})
\left[ \mathbf{g} \cdot (\nabla
  \mathbf{u}) \mathbf{n}\right]
+ \nonumber \\
& \mathbf{n} \otimes \mathbf{n}
\left[\text{Tr} (\nabla \mathbf{u}) - 2 \mathbf{g} \cdot (\nabla
  \mathbf{u}) \mathbf{g}\right], \\
\mathbf{a} & = \mathbf{g} \cdot \nabla \nabla(\mathbf{n} \cdot
\mathbf{u}).
\end{align}
Here, a tensor product $\mathbf{b} \otimes \mathbf{c}$ has components
$(\mathbf{b} \otimes \mathbf{c})_{ij} = b_i c_j$ and $a_i = g_k n_j
u_{j,ki}$, where repeated indices are summed according to the
Einstein convention.

We now specialize to linearization about a burning fixed point for
time-independent flows, and assume $\nabla | \mathbf{u} |^2 \ne 0$.
Equations~(\ref{r5}) -- (\ref{r9}) imply
\begin{equation}
  \mathsf{D} = \mathbf{n} \otimes \mathbf{n} \left[\text{Tr} (\nabla
    \mathbf{u}) - 2 \mu\right].
\end{equation}
Now, since $|\mathbf{g}| = 1$ is preserved by the dynamics, we need
only consider variations of $\mathbf{g}$ in the direction
$\mathbf{n}$; i.e. we consider the three-dimensional space of vectors
of the form $(\mathbf{v}, c \mathbf{n})$, with $\mathbf{v} \in
\Bbb{R}^2$ and $c \in \Bbb{R}$, which is invariant under $\mathsf{X}$.
Restricting $\mathsf{X}$ to this space yields the three-by-three
matrix
\begin{equation}
\mathsf{Y} = 
\left(
\begin{array}{cc}
(\nabla \mathbf{u})^T &  
-v_0 \mathbf{g}  \\
\mathbf{a}^T  &  
\text{Tr} (\nabla \mathbf{u}) - 2 \mu
\end{array}
\right).
\end{equation}
Equation~(\ref{r9}) implies that the two-dimensional space of vectors
of the form $(a \mathbf{g}, c )$, with $a, c \in \Bbb{R}$, is
invariant under $\mathsf{Y}$.  Restricting $\mathsf{Y}$ to this space
yields the two-by-two matrix
\begin{equation}
\mathsf{Z} = 
\left(
\begin{array}{cc}
\mu &  -v_0   \\
\mathbf{a} \cdot \mathbf{g}  &  
\text{Tr} (\nabla \mathbf{u}) - 2 \mu
\end{array}
\right).
\end{equation}
The eigenvalue of $\mathsf{Y}$ whose eigenvector does not have the
form $(a \mathbf{g}, c )$ is thus
\begin{equation}
\lambda_0 = \text{Tr}(\mathsf{Y}) - \text{Tr}(\mathsf{Z})  = \text{Tr}(\nabla
\mathbf{u}) - \mu
= \nu,
\end{equation}
where $\nu$ is the remaining (necessarily real) eigenvalue of $(\nabla
\mathbf{u})^T$.  The remaining two eigenvalues of $\mathsf{Y}$ are thus
the two eigenvalues of $\mathsf{Z}$ given by
\begin{equation}
\lambda_\pm = \frac{1}{2}\left( 
\nu \pm \sqrt{\nu^2 - 
4[v_0 \mathbf{a} \cdot \mathbf{g} + \mu(\nu - \mu)]}\right).
\label{r14}
\end{equation}

This expression can be put in a simpler form if we first consider the
components of $\mathbf{u}$ in the local frame of $\mathcal{C}$ (the
level set $|\mathbf{u}| = v_0)$,
defined by $\mathbf{e}_1'$ and $\mathbf{e}_2'$, the unit normal and
tangent vectors to $\mathcal{C}$, respectively.  The components of
$\mathbf{u}$ in the $(\mathbf{e}_1', \mathbf{e}_2')$ frame are denoted
$\mathbf{u}' = (u_1',u_2')$.  We parameterize $\mathcal{C}$ by the
Euclidean distance $\lambda$ measured along $\mathcal{C}$.  At a
burning fixed point, the derivative of $\mathbf{u}$ along
$\mathcal{C}$ is thus $d\mathbf{u}/d\lambda$, which by Eq.~(\ref{r9})
equals $\mu \mathbf{g}$, taking $\lambda$ to increase in the direction
$\mathbf{g}$.  The quantity $\mu/v_0 = \mu/|\mathbf{u}|$ is thus a
kind of rotation rate of $\mathbf{u}$ along $\mathcal{C}$.  Similarly,
we may differentiate $\mathbf{u}$ in the $(\mathbf{e}_1',
\mathbf{e}_2')$ frame, i.e. $d\mathbf{u'}/d\lambda$.  Since
$|\mathbf{u'}| = v_0$ is constant along $\mathcal{C}$,
$d\mathbf{u'}/d\lambda$ must point orthogonal to $\mathbf{u'}$.  At a
burning fixed point, $\mathbf{g}$ is orthogonal to $\mathbf{u}$, and
hence
\begin{equation}
d\mathbf{u'}/d\lambda = \mu' \mathbf{g'},
\label{r24}
\end{equation}
where $\mathbf{g'} = (0,1 )$ is the components of $\mathbf{g}$ in the
$(\mathbf{e}_1', \mathbf{e}_2')$ frame.  The (real) quantity $\mu'/v_0$ is thus
also the rotation rate of $\mathbf{u}$ along $\mathcal{C}$, but
viewed in the $(\mathbf{e}_1', \mathbf{e}_2')$ frame.  Appendix
\ref{sec:derivations} shows
\begin{equation}
\mu' = \mu + v_0 \kappa, 
\label{r12}
\end{equation}
where $\kappa$ is the (signed) curvature of $\mathcal{C}$, which equals (see
Appendix \ref{sec:derivations}) 
\begin{equation}
\kappa = \frac{v_0 \mathbf{a} \cdot \mathbf{g} - \mu^2}{v_0 \nu}
\label{r13}
\end{equation}
at a burning fixed point.  Combining Eqs.~(\ref{r12}) and (\ref{r13})
yields
\begin{equation}
\mu' = \frac{1}{\nu} \left[ v_0 \mathbf{a} \cdot \mathbf{g} + \mu
  (\nu -\mu) \right],
\label{r23}
\end{equation}
which can be used to transform Eq.~(\ref{r14}), as below.
\begin{theorem}
\label{t3}
For a time-independent flow $\mathbf{u}$, the eigenvalues about a
burning fixed point are
\begin{align}
\lambda_0 &= \nu, \\
\lambda_\pm &= \frac{1}{2}\left( 
\nu \pm \sqrt{\nu^2 - 4 \nu \mu'} \right),
\label{r15}
\end{align}
where $\nu$ is the eigenvalue of $(\nabla \mathbf{u})^T$ \emph{not} given
by Eq.~(\ref{r9}) and $\mu'$ is given by Eq.~(\ref{r24}).  The linear
stability of a burning fixed point is thus determined by the signs of
$\nu$ and $\mu'$ according to the following table.
\begin{center}
\begin{tabular}{l|cc}
& $\nu < 0$ & $\nu > 0$ \\
\hline
$\mu' > 0$ & SSU & UUU \\
$\mu' < 0$ & SSS & SUU 
\end{tabular}
\end{center}
\end{theorem}

In the above, $\nu$ determines the stability of two eigenvalues and
$\mu'$ the stability of one.  In either the SSU or SUU case, all
eigenvalues are real.  However, in the UUU or SSS case, two
eigenvalues are complex if $4 \mu'/\nu > 1$.  In
Sect.~\ref{BIMBarriers} we shall pay particular attention to the SSU
case.  For incompressible flows $\nu = -\mu$, and Eq.~(\ref{r23})
reduces to
\begin{equation}
\mu'  = \frac{1}{\mu} \left[ 2 \mu^2  - v_0 \mathbf{a} \cdot \mathbf{g} \right], 
\label{r16}
\end{equation}
and Theorem~\ref{t3} specializes to the following.
\begin{theorem}
\label{t4}
  For a time-independent, incompressible flow $\mathbf{u}$, the eigenvalues
  about a burning fixed point are
\begin{align}
\lambda_0 &= -\mu, \label{r27}\\
\lambda_\pm &= \frac{1}{2}\left( 
-\mu \pm \sqrt{\mu^2 + 4 \mu \mu'} \right), \label{r28}
\end{align}
where $\mu$ and $\mu'$ are given by Eqs.~(\ref{r9}) and (\ref{r24}).
The linear stability of a burning fixed point is thus determined by the signs
of $\mu$ and $\mu'$ according to the following table.
\begin{center}
\begin{tabular}{l|cc}
& $\mu > 0$ & $\mu < 0$ \\
\hline
$\mu' > 0$ & SSU & UUU \\
$\mu' < 0$ & SSS & SUU 
\end{tabular}
\end{center}
\end{theorem}

For a linear flow ($\mathbf{a} = 0$), Eqs.~(\ref{r27}) and (\ref{r28})
reduce to the eigenvalues in Table~\ref{Table:BFP}.  More generally,
for either sufficiently small $v_0$ or sufficiently linear flow
($\mathbf{a}$ small), Eq.~(\ref{r16}) shows that $\mu'$ has the same
sign as $\mu$, and thus a burning fixed point has one of only two
stability types: SSU ($\mu > 0$) or SUU ($\mu < 0$).  Referring to
Fig.~\ref{fig:ConDiFields}, defocusing flows yield SSU and focusing
flows SUU, as verified by the fixed points in
Fig.~\ref{fig:hyperbolic}a.

\subsection{Topological index theory for the existence of burning fixed
  points}

\label{sec:topindex}

\begin{figure}
\centering
\includegraphics[width=\linewidth]{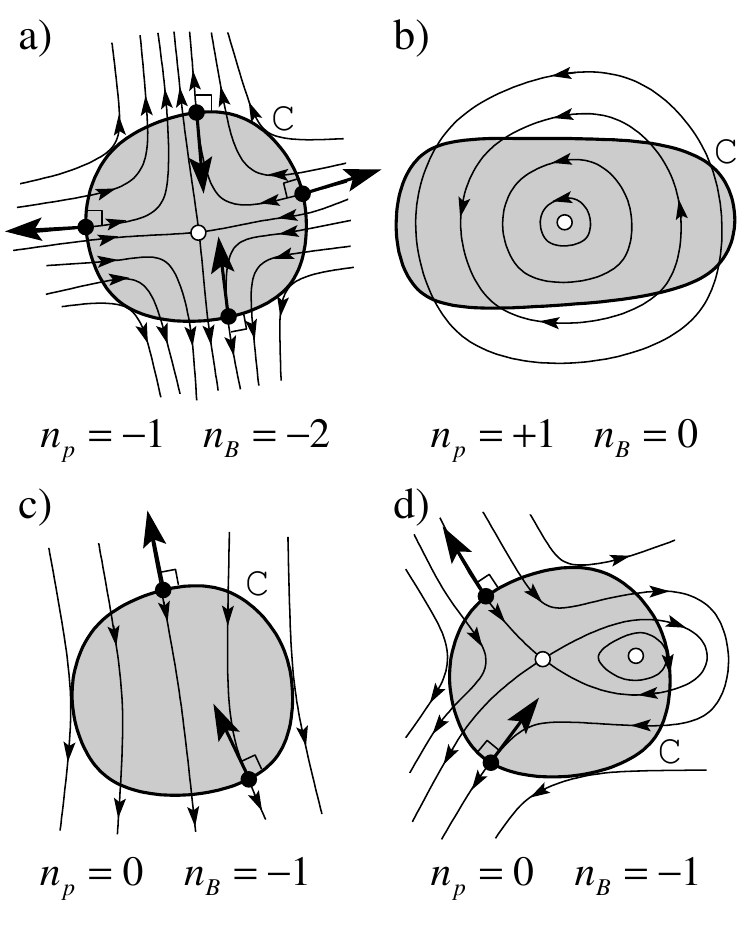}
\caption{Four flows illustrating Poincar\'{e} and burning indices. }
\label{fig:topindex}
\end{figure}

The number and stability type of burning fixed points are constrained
by a topological index theory similar to the Poincar\'{e} index theory
for fixed points in two-dimensional flows~\cite{Guckenheimer83}.
Following Theorem~\ref{t2}, we consider a closed loop $\mathcal{C}$
upon which $|\mathbf{u}| = v_0$.  Since $|\mathbf{u}| \ne 0$ on this
loop, the winding of the orientation of $\mathbf{u}$ can be used to
define two topological indices.  The first is the usual Poincar\'{e}
index~\cite{Guckenheimer83}.
\begin{equation}
  n_P(\mathcal{C}) = \parbox[t]{.80 \linewidth}{the number of
    counterclockwise (ccw) rotations made by $\mathbf{u}$
    under one ccw circuit of
    $\mathcal{C}$.}
\end{equation}
Fig.~\ref{fig:topindex} illustrates $n_P$ for four example flows.  The
reader is encouraged to confirm the value of $n_P$ cited in each case.
Following standard practice, a Poincar\'{e} index is also defined for
each advective fixed point $\mathbf{z}$ as follows.
\begin{equation}
   n_P(\mathbf{z}) = 
\begin{cases}
+1 & \mathbf{z} \text{ elliptic} \\
-1 & \mathbf{z} \text{ hyperbolic} 
\end{cases}
\end{equation}
The Poincar\'{e} index of the loop $\mathcal{C}$ is then equal to the
sum of the Poincar\'{e} indices of all the enclosed fixed points
\begin{equation}
n_P(\mathcal{C}) = \sum_{\text{fixed points } \mathbf{z}}
n_P(\mathbf{z})
= N_e - N_h,
\label{r36}
\end{equation}
where $N_e$ and $N_h$ are the number of elliptic and hyperbolic
advective fixed points, respectively.  The reader can confirm this
relationship in Fig.~\ref{fig:topindex}.

Alternatively, by considering the components $\mathbf{u}'$ in the
local frame $(\mathbf{e}_1', \mathbf{e}_2')$ of $\mathcal{C}$, we
define the \emph{burning index} as follows.
\begin{equation}
  n_B(\mathcal{C})  = \parbox[t]{.80 \linewidth}{the number of ccw rotations made by $\mathbf{u}'$
    under one ccw circuit of
    $\mathcal{C}$.}
\end{equation}
The two indices are related by
\begin{equation}
n_B(\mathcal{C})  = n_p(\mathcal{C})  - 1,
\label{r17}
\end{equation}
since the vector $\mathbf{u}$ in the original laboratory frame
undergoes one additional rotation relative to the vector $\mathbf{u}'$
in the local frame of $\mathcal{C}$.

\begin{figure}
\centering
\includegraphics[width=\linewidth]{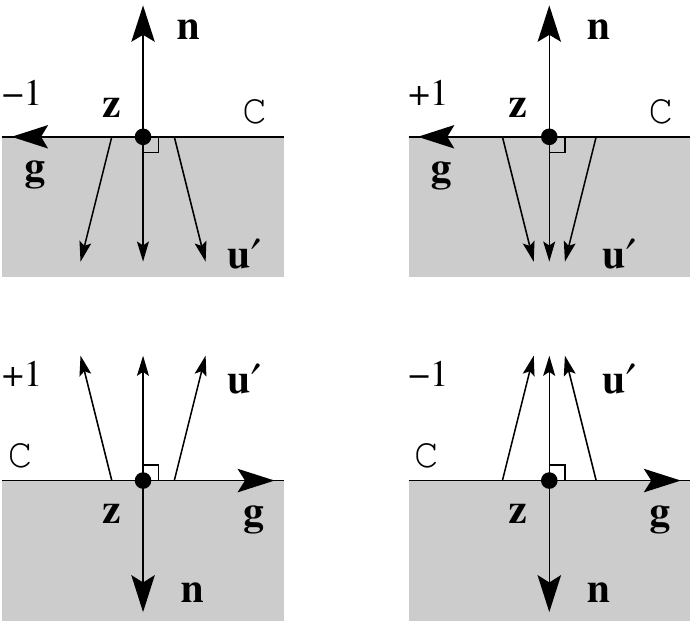}
\caption{The burning index $\pm 1$ is given for each velocity
  configuration.  The shaded region satisfies $|\mathbf{u}| < v_0$.}
\label{fig:ConDiPrimeFields}
\end{figure}

For each rotation made by $\mathbf{u}'$, $\mathbf{u}'$ is
perpendicular to $\mathcal{C}$ twice, once on either side of
$\mathcal{C}$.  Thus, there are at least $|n_B|$ outward-burning fixed
points on $\mathcal{C}$ and $|n_B|$ inward-burning fixed points.  This
result can be refined by defining a burning index for each burning
fixed point $\mathbf{z}$.
\begin{equation}
   n_B(\mathbf{z}) = 
\begin{cases}
  +1 & \mathbf{u'} \text{ rotates ccw at } \mathbf{z} \text{,
    traversing $\mathcal{C}$ ccw} \\
  -1 & \mathbf{u'} \text{ rotates cw at } \mathbf{z} \text{,
    traversing $\mathcal{C}$ ccw}
\end{cases}
\end{equation}
This definition is illustrated by the four cases in
Fig.~\ref{fig:ConDiPrimeFields}.  These four cases also make it clear
that the burning index of a fixed point can be re-expressed in terms
of $\mu'$ [Eq.~(\ref{r24})] as
\begin{equation}
   n_B(\mathbf{z}) = 
\begin{cases}
\left.
\begin{array}{ll}
+1 &   \mu'<0 \\
-1 &   \mu'>0 
\end{array}
\right\}
& \parbox{1.75in}{and  $\mathbf{u}(\mathbf{z})$ points inward to
  $\mathcal{C}$ (outward-burning)} \\
\\
\left.
\begin{array}{ll}
+1 &   \mu'>0 \\
-1 &   \mu'< 0 
\end{array}
\right\} & \parbox{1.75in}{and $\mathbf{u}(\mathbf{z})$ points outward
  to
  $\mathcal{C}$ (inward-burning)} \\
\end{cases}
\label{r25}
\end{equation} 

Analogous to the Poincar\'{e} index, the burning index
$n_B(\mathcal{C})$ of the curve $\mathcal{C}$ is the sum of the
burning indices of either all the outward-burning fixed points or all
the inward-burning fixed points located on $\mathcal{C}$.
\begin{align}
n_B(\mathcal{C}) &= \sum_{\text{out BFPs }\mathbf{z}}
n_B(\mathbf{z})
= N^\text{out}_- - N^\text{out}_+ \\
&= \sum_{\text{in BFPs }\mathbf{z}}
n_B(\mathbf{z})
= N^\text{in}_+ - N^\text{in}_-,
\end{align}
where $N^\text{out}_+$ and $N^\text{out}_-$ are the number of
outward-burning fixed points with $\mu' > 0$ and $\mu' < 0$,
respectively, and $N^\text{in}_+$ and $N^\text{in}_-$ are the number
of inward-burning fixed points with $\mu' > 0$ and $\mu' < 0$,
respectively.  Finally, the number and type of burning fixed points on
$\mathcal{C}$ can be directly related to the advective fixed points
interior to $\mathcal{C}$ through Eqs.~(\ref{r36}) and (\ref{r17}).
\begin{equation}
  N^\text{out}_--N^\text{out}_+ = N^\text{in}_+ - N^\text{in}_- = N_e -
  N_h - 1.
\label{r26}
\end{equation}

For an incompressible flow, $\mu$ must be positive when $\mathbf{u}$
points inward and negative when $\mathbf{u}$ points outward.  Using
the table in Theorem~\ref{t4}, Eq.~(\ref{r25}) then becomes
\begin{equation}
   n_B(\mathbf{z}) = 
\begin{cases}
\begin{array}{ll}
+1 &  \text{stability } SSS \text{ and outward-burning} \\
-1 &  \text{stability } SSU  \text{ and outward-burning} \\
\\
+1 &   \text{stability } UUU \text{ and inward-burning} \\
-1 &   \text{stability } SUU \text{ and inward-burning} 
\end{array}
\end{cases}
\end{equation} 
Similarly, Eq.~(\ref{r26}) becomes
\begin{equation}
N^\text{out}_{SSS}-N^\text{out}_{SSU} = N^\text{in}_{UUU} - N^\text{in}_{SUU} = N_e - N_h - 1,
\end{equation}
where $N^\text{in/out}_{XXX}$ measures the number of
inward/outward-burning fixed points of the given stability type.  This
equation is confirmed in Fig.~\ref{fig:topindex}, where all burning
fixed points either have stability $SSU$ or $SUU$ and hence the number
of outward- or inward-burning fixed points simply equals $1 - N_e +
N_h = -n_B(\mathcal{C})$.

\section{Burning invariant manifolds (BIMs): one-sided barriers to
  front propagation}

\label{BIMBarriers}

Having examined the existence and stability of burning fixed points in
Sect.~\ref{sec:BFP}, we now focus on the invariant manifolds attached
to them.  These manifolds are invariant either under the differential
equation (\ref{r3}) [or (\ref{eq:3DODE})], in the time-independent
case, or under the Poincar\'{e} map, in the time-periodic case; a
stable/unstable invariant manifold consists of all points that
converge upon a burning fixed point in forward/backward
time~\cite{Guckenheimer83}.  As noted above, we call these stable and
unstable manifolds \emph{burning} invariant manifolds (BIMs) to
distinguish them from the traditional invariant manifolds for
advection.  Due to the special role played by one-dimensional
invariant manifolds, all BIMs subsequently discussed are assumed to be
one-dimensional, unstable manifolds, attached to a fixed point of
stability SSU, as shown in Figs.~\ref{fig:tindepBIMs} and
\ref{fig:tdepBIMs} for time-independent and time-periodic flows.
(One-dimensional, stable BIMs are similarly analyzed by running time
backwards; they will be relevant in a future publication for
developing a theory of lobe dynamics for front propagation.)  The
central point of our theoretical development is that these BIMs form
one-sided barriers to front propagation (Figs.~\ref{fig:tindepBIMs}
and ~\ref{fig:tdepBIMs}).

\subsection{BIMs are fronts}

\label{sec:BIMfronts}

Though most one-dimensional curves in the three-dimensional
$xy\theta$-space are not fronts, an unstable (one-dimensional) BIM
attached to an SSU fixed point \emph{does} satisfy the front
compatibility criterion, i.e.  Eq.~(\ref{r4}) or (\ref{r6}), as
established here.  Consider first a short line segment $(x_*, y_*,
\theta_*) + (\lambda \cos(\theta), \lambda \sin(\theta), 0 )$,
parameterized by $-\epsilon < \lambda < \epsilon$ and passing through
a burning fixed point $(x_*,y_*,\theta_*)$ of stability SSU.  This
line clearly satisfies Eq.~(\ref{r4}).  Furthermore, due to the two
stable dimensions of the fixed point, as this curve evolves in time,
it converges upon the BIM, approximating progressively longer segments
of the BIM~
\cite{footnote3}
.  Since the front compatibility criterion is
preserved over time, the evolving curve remains a front, and hence the
BIM to which it converges satisfies the front compatibility criterion
everywhere along its length.  This argument applies for both
time-independent and time-periodic flows.

\subsection{BIMs are one-sided barriers}

\label{sec:onesided}

As pointed out in Figs.~\ref{fig:tindepBIMs} and \ref{fig:tdepBIMs},
BIMs are one-sided barriers, preventing fronts from passing in one
direction, but not the other.  We can now explain this fact as
follows.  Since BIMs are fronts, they satisfy the no-passing lemma
(Sect.~\ref{sec:nopassing}).  The local formulation of this lemma then
implies that no impinging front can burn past any point of the BIM in
the same direction that the BIM itself is burning.

In some flows, a BIM can partition the fluid into two distinct
regions, one entirely burned and one entirely unburned.  The BIM then
prevents the burned region from penetrating the unburned region.  This
is seen, for example, in the hyperbolic flow in
Fig.~\ref{fig:hyperbolic}a, in which the BIMs form two vertical lines.
If the entire vertical strip between the two BIMs were burned, the
region outside the strip would remain unburned forever; consequently,
a reaction catalyzed anywhere between the BIMs will never burn outside
the central strip.  Alternatively, the BIMs in
Figs.~\ref{fig:tindepBIMs} and \ref{fig:tdepBIMs} do not confine the
burned material to a bounded region forever; in these examples, fronts
will eventually reach the other side of any given BIM.  This can
happen in two ways, discussed below.

\subsection{Cusps form openings in the barrier}
\label{sec:cusps}

\begin{figure}
\centering
\includegraphics[width=\linewidth]{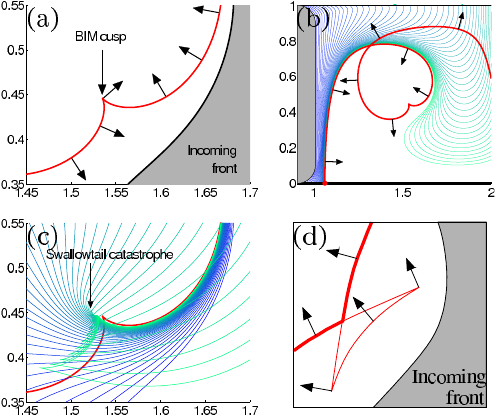}
\caption{(color online) 
  a) A BIM cusp. 
  b) Zoom out of cusp with fronts wrapping around. 
  c) Fronts spiraling and sliding past the cusp. 
  d) A BIM with swallowtail structure.}
\label{fig:cusp}
\end{figure}

The $xy$-projection of a BIM may form cusps, as seen in
Figs.~\ref{fig:tindepBIMs} and \ref{fig:tdepBIMs}.  A cusp in a BIM
has a profound impact on the bounding nature of the BIM.  This is
because the burning direction (and hence bounding direction) of a BIM
is reversed at a cusp, relative to an impinging front.  See
Fig.~\ref{fig:cusp}a.  To the right of the cusp, the BIM is burning
toward the upper left, acting as a barrier to the impinging front.
However, left of the cusp, the BIM is burning toward the lower right.
This flip in the burning direction results from the continuity of
$\theta$ along the curve in the full $xy\theta$-space---note that the
arrows in Fig.~\ref{fig:cusp}a rotate smoothly along the BIM.  Thus,
left of the cusp, the BIM does not act as a barrier to the front
below; the front will simply pass through the left segment, as in
Fig.~\ref{fig:cusp}c.  Left of the cusp, the front \emph{spirals}
clockwise around the cusp and forms a swallowtail singularity.  The
front then \emph{slides} past the cusp itself, moving toward the lower
right and crossing over the right BIM segment from above.  The
critical observation is that the cusp provides an opening, through
which the front can pass around the bounding segment of the BIM, and
then collide with it.  For time-independent flows, this is the only
way for a front to reach the other side of a BIM.

Figure~\ref{fig:cusp}b shows a zoomed out view of the BIM and an
initial front impinging from the left.  Note that the evolving front
passes right through the continuation of the BIM segment shown left of
the cusp in Fig.~\ref{fig:cusp}a, owing to the orientation of its
burning direction.  However, although a cusp creates an opening in the
bounding nature of a BIM, the entire remainder of the BIM is not
necessarily irrelevant.  Figure~\ref{fig:cusp}d shows an example of
how a second cusp again reverses the burning direction
\cite{footnote4}.
%
Though the front may burn around each of the two cusps, the (bold)
intervals still form a barrier to further front propagation.

\subsection{Moving past BIMs in time-periodic flows}

\label{sec:movepast}

\begin{figure}
\centering
\includegraphics[width=1\linewidth]{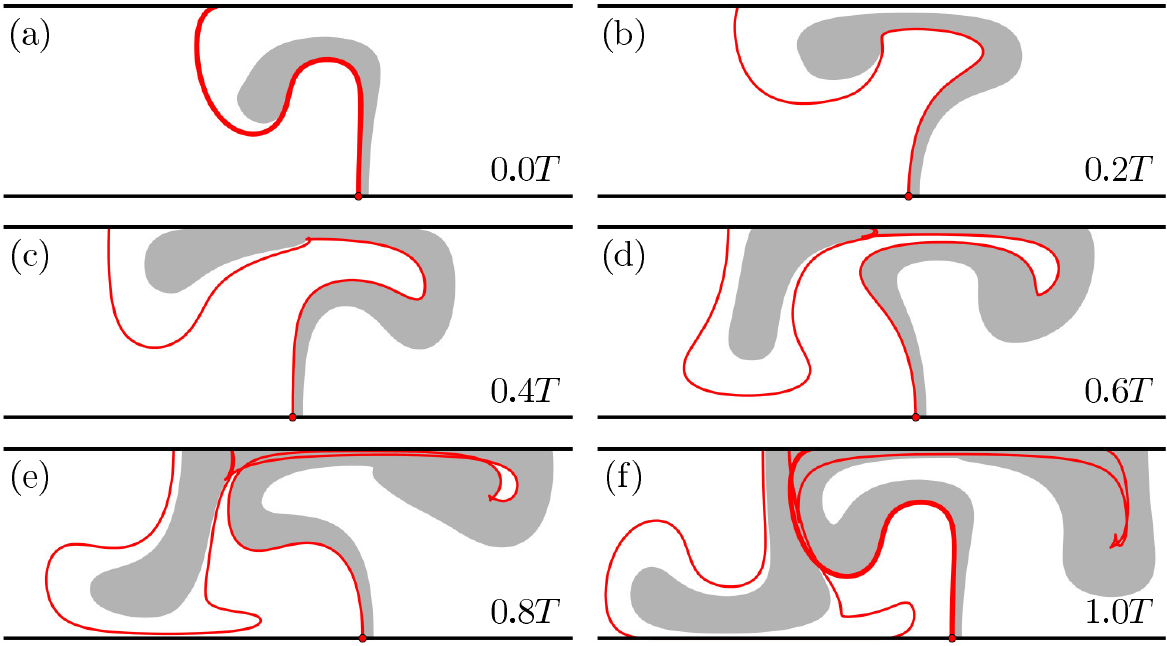}
\caption{(color online) The mechanism by which a burned region (gray)
  may move past an initial BIM segment (red) under time-periodic
  driving.}
\label{fig:evolution}
\end{figure}

For a time-periodic flow, a front may reach the other side of a BIM by
a second mechanism illustrated in Fig.~\ref{fig:evolution}.
Figure~\ref{fig:evolution}a shows an initial segment of a left-burning
BIM (bold, red) that transects the fluid channel.  A gray burned region is
immediately right of the BIM.  As the system evolves over one driving
period, the BIM segment stretches and folds, ultimately crossing over
itself (in the $xy$-projection).  By the final frame, both the BIM and
the front extend left of the original BIM segment (bold).  While this
may appear to violate the no-passing lemma, at no point does the
burned region cross over any point of the BIM in the burning direction
of the BIM.  Furthermore, at each time step, the leftmost transecting
segment of the BIM bounds the burned region.  By the final frame, the
burned region has moved past the original BIM segment because of the
stretching and folding of the BIM itself.  This mechanism is analogous
to that of lobe dynamics (or turnstiles) for the advective transport
of passive impurities in time-periodic
flows~\cite{MacKay84,Rom-Kedar90b,Wiggins92,Solomon96}.

\subsection{Convergence to BIMs and experimental measurement}

\label{sec:convergence}

Fronts are not only bounded by BIMs but \emph{converge} upon them.
This is due to the transverse stability of a BIM in the neighborhood
of its SSU fixed point (a fact already utilized in
Sec.~\ref{sec:BIMfronts}).  If a front has one front element that
converges on a burning fixed point
\cite{footnote5},
%
nearby front elements will evolve arbitrarily close to any given point of
the BIM over time.  This convergence is clearly witnessed in
Figs.~\ref{fig:tindepBIMs} and \ref{fig:tdepBIMs}b, d.

Front convergence suggests a protocol to measure BIMs in the
laboratory.  For the time-independent case, first catalyze a reaction
front sufficiently near the burning fixed point.  (The advective fixed
point is often a good choice.)  Then observe the limiting behavior of
the evolving front, which identifies the BIM location, as simulated in
Fig.~\ref{fig:tindepBIMs}a.  Points on the BIM nearest the fixed point
will approach their limiting value earlier than points further along
the BIM, and hence can be measured with greater accuracy
\cite{footnote6}.
%
This protocol has been successfully implemented in the Solomon lab, as
discussed in the companion article \cite{Bargteil12} and
\setcitestyle{numbers} Ref.~\cite{Mahoney12}.  

For the time-periodic case, a similar protocol exists, with one
critical change.  Care must be taken to record the converging front at
a constant phase of the driving (say $t = 0$).  Hence, multiple
experimental runs are needed in which the reaction is stimulated at
incrementally earlier times $t < 0$, and then recorded at the common
time $t = 0$.  See Ref. \cite{Mahoney12} for further details and an
experimental implementation.  

BIMs can also be computed directly from the dynamics
Eq.~(\ref{eq:3DODE}) if the flow $\mathbf{u}(\mathbf{r},t)$ is known.
However, the ability to measure BIMs directly in experimental flows
frees one from the requirement of either theoretically modeling
$\mathbf{u}(\mathbf{r},t)$ or experimentally measuring
$\mathbf{u}(\mathbf{r},t)$, e.g. from particle-tracking data.
\setcitestyle{super}

As a front is converging upon a given point $\mathbf{r}$ of a BIM,
another interval of the front may have taken a different route to
collide with the BIM at $\mathbf{r}$ from the other side (as discussed
in Sects.~\ref{sec:cusps} and \ref{sec:movepast}).  Once this happens,
one can no longer image the convergence upon $\mathbf{r}$, since the
entire neighborhood of $\mathbf{r}$ is burned.  Thus, the accuracy by
which the BIM can be measured is limited by the time it takes another
interval of the front to collide with the BIM.  The best strategy then
is to initialize the front as close to the burning fixed point as
possible, allowing the most time to approach any given point on the
BIM before the convergence process is wiped out.  An enhanced
strategy, permitting a more detailed measurement of the BIMs, would be
to also suppress, or ``erase'', the second front before it collides
with the point $\mathbf{r}$ being measured.  This could be done
optically for certain photosensitive reactions.

\begin{table}[t]
\underline{\makebox[\linewidth][c]{Front Propagation Effects} }
\begin{enumerate}[$\bullet$]
\item Growth in the total area of the impurity region
\item   Formation of cusps, swallowtails, and self-intersections at the
  impurity boundary (and within a BIM)
\item Directionality (i.e. existence of a burning direction) for the
  impurity boundary (and for a BIM)
\end{enumerate}
\underline{\makebox[\linewidth][c]{Advection Effects} }
\begin{enumerate}[$\bullet$]
\item   Existence of fixed points and invariant manifolds (BIMs) 
\item Bounding property of invariant manifolds and use in transport
\item Convergence of impurity boundary to invariant manifolds
\item Lobe dynamics approach to transport of impurity region
\end{enumerate}
\caption{\label{table2} Comparison of front versus advection effects}
\end{table}

\section{Concluding Remarks}

\label{sec:limits}

A useful framework for organizing the properties of front propagation
in a flowing medium [Eq.~(\ref{r3})] is to consider the two natural
limiting cases: $\mathbf{u} = 0$ (front propagation in a stationary
medium) and $v_0 = 0$ (passive advection).  In both limits, one is
concerned with the change in geometry of an ``impurity'' region
(either passive tracers or an active chemical medium) within a
``background'' medium.  The front dynamics, and the BIMs in
particular, reflect properties unique to one limit or the other, as
summarized in Table~\ref{table2}.  For example, the formation of cusps
in a BIM reflects behavior typical in front propagation, but does not
occur for invariant manifolds used in advection.  Similarly, BIMs have
a burning direction, which is natural for fronts, but absent for
advective manifolds.  On the other hand, the very existence of BIMs,
their use as bounding curves, and their application to transport
reflect similarities with the advection problem but are absent from
studies of front propagation in homogeneous stationary media.

\acknowledgments

This work was supported by the US National Science Foundation under
grant PHY-0748828.  The authors gratefully acknowledge extensive
discussion with Tom Solomon and Dylan Bargteil.

\appendix
\section{Derivations}

\label{sec:derivations}

\subsection{Derivation of Eq.~(\ref{r12})}

The (signed) curvature of a planar curve $\mathcal{C}$ equals
\begin{equation}
\kappa = \mathbf{e}_2' \cdot \frac{d \mathbf{e}_1'}{d \lambda},
\label{r32}
\end{equation}
where $\mathbf{e}_1'$ and $\mathbf{e}_2'$ are the normal and tangent
unit vectors to the curve, as in Sec.~\ref{sec:stability}.  (Here,
$\kappa < 0$ implies $\mathbf{e}_1'$ points toward the center of
curvature.)  We denote $\mathsf{R}(\lambda)$ as the rotation matrix
connecting the $(\mathbf{e}_1',\mathbf{e}_2')$ frame to the laboratory
frame $(\mathbf{e}_1,\mathbf{e}_2) =
(\mathbf{\hat{x}},\mathbf{\hat{y}})$, i.e.
\begin{equation}
\mathbf{e}_i'(\lambda) = \mathsf{R}(\lambda) \mathbf{e}_i, \; i =
1,2.
\label{r35}
\end{equation}
The \emph{components} of a vector then transform using $\mathsf{R}^T$,
e.g.
\begin{equation}
\mathbf{u}'(\lambda) = \mathsf{R}^T(\lambda) \mathbf{u}(\lambda). \label{r33}
\end{equation}
Specializing to a burning fixed point, at which $\mathbf{n} =
\mathbf{e}_1'$ and $\mathbf{g} = \mathbf{e}_2'$ (Theorem~\ref{t2}), Eqs.~(\ref{r32}) and (\ref{r35}) yield
\begin{equation}
\kappa = \mathbf{e}_2' \cdot \frac{d \mathsf{R}}{d \lambda}
\mathsf{R}^T \mathbf{e}_1'
= \mathbf{g} \cdot \frac{d \mathsf{R}}{d \lambda} \mathsf{R}^T \mathbf{n}.
\label{r34}
\end{equation}
From $d \mathbf{u}/d \lambda = \mu \mathbf{g}$, we then have
\begin{align}
\mu  &= \mathbf{g} \cdot \frac{d \mathbf u}{d \lambda} 
= \mathbf{g} \cdot \left[\frac{d \mathsf{R}}{d \lambda} \mathsf{R}^T
\mathbf{u}
+ \mathsf{R}\frac{d \mathbf{u}'}{d \lambda} \right] \nonumber \\
&= \mathbf{g} \cdot \left[-v_0 \frac{d \mathsf{R}}{d \lambda} \mathsf{R}^T
\mathbf{n}
+ \mu' \mathsf{R}\mathbf{g}' \right]  
=-v_0 \kappa + \mu', 
\end{align}
where the second equality follows from Eq.~(\ref{r33}), the third from
Eqs.~(\ref{r5}) and (\ref{r24}), and the last from Eqs.~(\ref{r34})
and the fact $\mathbf{g} = \mathsf{R}\mathbf{g}'$.  This yields Eq.~(\ref{r12}).

\subsection{Derivation of Eq.~(\ref{r13})}

The (signed) curvature of the level set $f(\mathbf{r}) = 0$ for a
general function $f$ is
\begin{equation}
\kappa = \frac{\mathbf{e}_2' \cdot (\nabla
\nabla f) \mathbf{e}_2'}{\nabla f \cdot \mathbf{e}_1'}. 
\label{r29}
\end{equation}
Taking $f = |\mathbf{u}|^2 - v_0^2$ defines the level set
$\mathcal{C}$.  Since $\mathbf{e}_1' = \mathbf{n}$ and $\mathbf{e}_2' =
\mathbf{g}$ at a burning fixed point, the denominator of
Eq.~(\ref{r29}) becomes
\begin{align}
(\nabla |\mathbf{u}|^2) \cdot \mathbf{n} &= -2 v_0 \mathbf{n} \cdot
(\nabla \mathbf{u}) \mathbf{n} \nonumber \\ 
&= -2 v_0 ( \mbox{Tr} (\nabla \mathbf{u}) -  
\mathbf{g} \cdot
(\nabla \mathbf{u}) \mathbf{g} )= -2 v_0 \nu,
\label{r30}
\end{align}
where the first equality follows from Eq.~(\ref{r5}) and the final
equality from Eq.~(\ref{r9}) and the fact that $\mbox{Tr} (\nabla
\mathbf{u}) = \mu + \nu$.  Similarly, a straightforward application of
Eqs.~(\ref{r5}) and (\ref{r9}) reveals the numerator of
Eq.~(\ref{r29}) to be
\begin{align}
  \mathbf{g} \cdot (\nabla \nabla |\mathbf{u}|^2 ) \mathbf{g}
= 2 \mu^2 - 2v_0 \mathbf{a}\cdot\mathbf{g}.
\label{r31}
\end{align}
Eq.~(\ref{r13}) then follows from Eqs.~(\ref{r29}) -- (\ref{r31}).


\begin{thebibliography}{51}
\bibitem [{Note1()}]{footnote1}%
  \BibitemOpen
  \bibinfo {note} {Throughout, we use the term ``burned'' or ``burning''
  generically for front propagation}\BibitemShut {NoStop}%
\bibitem [{Note3()}]{footnote3}%
  \BibitemOpen
  \bibinfo {note} {If the original segment were to lie exactly within the
  two-dimensional stable manifold of the fixed point, the segment would
  converge to the fixed point only. In this case, the original segment could be
  perturbed slightly out of the stable manifold everywhere except the fixed
  point, while still preserving Eq.~(\ref {r4}).}\BibitemShut {Stop}%
\bibitem [{Note4()}]{footnote4}%
  \BibitemOpen
  \bibinfo {note} {The geometry of Fig.~\ref {fig:cusp}d can only happen in the
  time-periodic case. In the time-independent case, it is forbidden because:
  (i) the BIM consists of a single trajectory and (ii) this trajectory would
  violate the no-passing lemma if it were to pass through the intersection
  point of the swallowtail in Fig.~\ref {fig:cusp}d.}\BibitemShut {Stop}%
\bibitem [{Note5()}]{footnote5}%
  \BibitemOpen
  \bibinfo {note} {Equivalently, we may consider any front that intersects the
  \protect \emph {stable} manifold of the fixed point, or alternatively, any
  front for which the stable manifold passes through the burned
  region.}\BibitemShut {Stop}%
\end{thebibliography}

%

\end{document}